\documentclass[article,prl,balancelastpage,twocolumn,showpacs]{revtex4}
\usepackage{makeidx}
\usepackage{amssymb}
\usepackage{dcolumn}
\usepackage{graphicx}
\usepackage{acronym}

\input{tcilatex}

\begin{document}

\title{Spin oscillations of relativistic fermions in the field of a
traveling circularly polarized electromagnetic wave and a constant magnetic
field}
\author{B. V. Gisin }
\affiliation{IPO, Ha-Tannaim St. 9, Tel-Aviv 69209, Israel. E-mail:
borisg2011@bezeqint.net}
\date{}

\begin{abstract}
The Dirac equation, in the field of a traveling circularly polarized
electromagnetic wave and a constant magnetic field, has singular solutions,
corresponding the expansion of energy in vicinity of some singular point.
These solutions described relativistic fermions. States relating to these
solutions are not stationary. The temporal change of average energy,
momentum and spin for single and mixed states is studied in the paper. A
distinctive feature of the states is the disappearance of the longitudinal
component of the average spin. Another feature is the equivalence of the
condition of fermion minimal energy and the classical condition of the
magnetic resonance.

Finding such solutions assumes the use of a transformation for rotating and
co-moving frames of references. Comparison studies of solutions obtained
with the Galilean and non-Galilean transformation shown that some parameters
of the non-Galilean transformation may be measured in high-energy physics.
\end{abstract}

\pacs{03.65.Ge, 71.70.Di, 13.49.Em; \hspace{20cm}}
\maketitle

\section{Introduction}

Recently, a new class of exact localized non-stationary solutions of the
Dirac equation in the field of a traveling circularly polarized
electromagnetic wave and a constant magnetic field was presented \cite{S1}.
These solutions correspond to stationary states (Landau levels \cite{Lqm})
in a rotating and co-moving frame of references. In contrast to classical
case, the wave function of such solutions never can be presented as large
and small two-component spinor. They are relevant only for relativistic
fermions.

Solutions, corresponding to the expansion of energy in vicinity of some
singular point, or singular solutions, are of interest from the experimental
viewpoint. Since solutions are non-stationary, particular attention in the
paper is given to the study of the temporal behavior of the average energy,
momentum and spin.

In the paper the Galilean transformation is used for the transition to the
rotating and co-moving frame of references, the term "rotating frame"
relates below to the given case. The term "initial frame" relates to the
resting (laboratory) frame.

But the issue arises what should be the transformation Galilean or not? It
means, is time the same in the rotating frame and\ the initial?

Some theoretical and experimental aspects of the problem in application to
optics are presented in \cite{S2}. Simple two-dimensional transformation $%
\tilde{\varphi}=\varphi -\Omega t,$ \ $\tilde{t}=-\tau \varphi +t$ was
discussed in this paper as an example. Existing experimental data in optics 
\cite{jps} allow to determine that the upper boundary of the parameter $\tau 
$ is$\ \sim 10^{-23}\sec $ \cite{S2}. In terms of length this corresponds to
a distance of the order of the proton size.

Unfortunately, the parameter $\tau $ vanishes from final results, for
bounded and square integrable solutions of Dirac's equation in the case the
two-dimensional transformation, after the mandatory reverse transformation
to the initial frame.

Comparison studies of solutions corresponding to a general form of the
three-dimensional non-Galilean transformation is performed in the paper.
This transformation not only is considered as a convenient tool for finding
solutions, but also as having a more general meaning.

\section{Solutions of Dirac's equation}

We consider Dirac's equation\ 
\begin{equation}
i\hbar \frac{\partial }{\partial t}\Psi =c\mathbf{\alpha }(\mathbf{\mathbf{p}%
}-\frac{e}{c}\mathbf{\mathbf{A}})\Psi +\beta mc^{2}\Psi =0  \label{Dir}
\end{equation}%
in the electromagnetic field with potential 
\begin{eqnarray*}
A_{1} &=&-\frac{1}{2}H_{3}y+\frac{1}{k}H\cos (\Omega t-kz), \\
A_{2} &=&\frac{1}{2}H_{3}x+\frac{1}{k}H\sin (\Omega t-kz),
\end{eqnarray*}%
where $k=\varepsilon \Omega /c$ is the propagation constant, $\Omega $ is
the frequency, the sign change of $\Omega $ corresponds to the opposite
polarization, values $\varepsilon =1$ and $\varepsilon =-1$ are used when
the wave propagates along the $z$-axis and opposite direction, respectively, 
$c$ is the speed of light, $\alpha _{k},\beta $ are Dirac's matrices, $H$ is
the amplitude of this wave. This potential corresponds to a traveling plane
circularly polarized wave and constant magnetic field.

By symmetry, the consideration of solutions is more convenient in a rotating
frame with coordinates%
\begin{eqnarray}
\tilde{x} &=&r\cos \tilde{\varphi},\text{ \ }\tilde{y}=r\sin \tilde{\varphi}%
,\   \label{xy} \\
\tilde{\varphi} &=&\varphi -\Omega t+kz,\text{ }\tilde{t}=t,\text{ }\tilde{z}%
=z,  \label{xyt}
\end{eqnarray}%
where the tilde corresponds to this frame.

Dirac's equation (\ref{Dir}) in the frame has exact localized stationary
solutions%
\begin{equation}
\tilde{\Psi}=\exp [-i\frac{\tilde{E}\tilde{t}}{\hbar }+i\frac{\tilde{p}%
\tilde{z}}{\hbar }-\frac{1}{2}d(\tilde{x}^{2}+\tilde{y}^{2})+d_{1}\tilde{x}%
+d_{2}\tilde{y}]\psi ,  \label{Psi}
\end{equation}%
were $\tilde{E}$ and $\tilde{p}$ is the "energy" and "momentum" along the $z$%
-axis", $\psi $ is a spinor.\ The equation with such a wave function may be
interpreted as a two-dimensional harmonic oscillator. The wave function in
the initial frame is $\Psi =\exp [-\frac{1}{2}\alpha _{1}\alpha _{2}(\Omega
t-kz)]\tilde{\Psi}$.

For all states parameters $d$ and $d_{1,}d_{2}$ are determined with help of
following relations%
\begin{equation}
d=\pm \frac{eH_{3}}{2\hbar c}>0,\text{ }d_{1}=-id_{2},\text{ }d_{2}=\frac{%
eHcd/k}{(\tilde{E}-c\tilde{p}\varepsilon )\Omega -2c\hbar dc},  \label{d012}
\end{equation}%
$d$ defines the characteristic size of the localization $l_{c}=\sqrt{%
\left\vert 2\hbar c/eH_{3}\right\vert }$. For simplicity, we consider below
the case $eH_{3}<0$.

States may be classified in accordance with the form of the spinor $\psi $.
A constant spinor describes "ground state". A spinor polynomial in $\tilde{x}
$,\ $\tilde{y}$ corresponds to "excited states".

\subsection{Ground state}

The solution for the ground state $\psi _{g}$ and the characteristic
equation, for energy eigenvalues $\mathcal{E}$, in normalized units are as
follows 
\begin{eqnarray}
\psi _{g} &=&N_{g}\psi _{0},\text{ \ }\psi _{0}=\left( 
\begin{array}{c}
h\mathcal{E} \\ 
-\varepsilon (\mathcal{E}+1)(\mathcal{E}-\mathcal{E}_{0}) \\ 
\varepsilon h\mathcal{E} \\ 
-(\mathcal{E}-1)(\mathcal{E}-\mathcal{E}_{0})%
\end{array}%
\right) ,\text{ \ }  \label{Sg} \\
\Lambda &=&\frac{2\varepsilon \tilde{p}c-\hbar \Omega }{mc^{2}},\text{ \ }%
\mathcal{E}(\mathcal{E}+\Lambda )-1-\frac{\mathcal{E}h^{2}}{\mathcal{E}-%
\mathcal{E}_{0}}=0,\text{ \ }  \label{Eqg}
\end{eqnarray}%
where normalized units are 
\begin{equation}
\mathcal{E}\equiv \frac{(\tilde{E}-\varepsilon \tilde{p}c)}{mc^{2}},\text{ \ 
}\mathcal{E}_{0}=\frac{2d\hbar }{\Omega m},\text{ \ }h=\frac{e}{kmc^{2}}H,
\label{Ng}
\end{equation}%
It is easy to prove the equality 
\begin{equation}
\int \tilde{\Psi}^{\ast }\tilde{\Psi}d\tilde{x}d\tilde{y}\equiv \int \Psi
^{\ast }\Psi dxdy=1,  \label{psps}
\end{equation}
for the rotating frame and the initial. The normalized constant $N_{g}$, as
usually, is defined from this equality.

For the $d=+eH_{3}/2\hbar c,$ the wave function is$\ \psi _{+}=\varepsilon
\alpha _{1}\alpha _{3}\beta \psi $ with the simultaneous sign change of $%
\mathcal{E}_{0}$.

\subsection{Excite states}

For the first "excite state" $\psi =\psi _{0}+\tilde{x}\psi _{x}+\tilde{y}%
\psi _{y}$, where $\psi _{0},\psi _{x},\psi _{y}$ are constant spinors. Two
types of such solutions are possible.

The first type is realized at $\psi _{y}=-i\psi _{x}.$ The solution and the
parameter $\Lambda $ of the characteristic equation are 
\begin{equation}
\psi _{e1}=N_{e1}\psi _{0}(1-i\frac{d}{d_{2}}\tilde{x}-\frac{d}{d_{2}}\tilde{%
y}),\text{ \ }\Lambda =\frac{2\varepsilon \tilde{p}c-3\hbar \Omega }{mc^{2}},
\label{we1}
\end{equation}%
where $\psi _{0}$ is the same as in (\ref{Sg}).

For the second type $\psi _{y}=i\psi _{x}.$The solution is

\begin{equation}
\psi =N_{e2}\left( 
\begin{array}{c}
(\mathcal{EE}_{0}+1)(\mathcal{E-E}_{0})-\varepsilon hk\mathcal{EE}_{0}(i%
\tilde{x}-\tilde{y}) \\ 
-\varepsilon (\mathcal{E}+1)\mathcal{E}_{0}[h-\varepsilon k(\mathcal{E}-%
\mathcal{E}_{0})(i\tilde{x}-\tilde{y})] \\ 
\varepsilon (\mathcal{EE}_{0}+1)(\mathcal{E-E}_{0})-hk\mathcal{EE}_{0}(i%
\tilde{x}-\tilde{y}) \\ 
-(\mathcal{E}-1)\mathcal{E}_{0}[h-\varepsilon k(\mathcal{E}-\mathcal{E}%
_{0})(i\tilde{x}-\tilde{y})]%
\end{array}%
\right) .  \label{Se2}
\end{equation}%
The \allowbreak eigenvalue equation differs from (\ref{Eqg}) by the
parameter $\Lambda =(2\varepsilon \tilde{p}c+\hbar \Omega )/mc^{2}$ and the
additional term $2\hbar \Omega \mathcal{E}_{0}/mc^{2}$ 
\begin{equation}
\mathcal{E}(\mathcal{E}+\Lambda )-1-\frac{2\hbar \Omega }{mc^{2}}\mathcal{E}%
_{0}-\frac{\mathcal{E}h^{2}}{\mathcal{E-E}_{0}}=0.  \label{Eqe1}
\end{equation}

Obviously, wave functions (\ref{Sg}), (\ref{Se2}) cannot be presented as a
small and large two-component spinor. It means that the difference $%
E^{2}-m^{2}c^{2}$ cannot be small and these solutions correspond only to the
relativistic case.

\section{Singular solutions}

The above non-stationary states can be determined with help of the "energy $%
\mathcal{E}$" in the rotating frame. However, in the initial frame, we
should use the average values {}{}of the operators. The same is valid for
momentum and spin. From the experimental point of view, of particular
interest are special solutions with energies close to the singular point $%
\mathcal{E}_{0}$. Such singular states for the Galilean transformation arise
only by a fixed momentum and relate as to the ground as exited states.

It is well known that the eigenvalue equation, as an equation of the third
order, has exact solutions $\mathcal{E}(\mathcal{E}_{0},h,\tilde{p})$.
However, for singular solutions it is more convenient to use roots of the
eigenvalue equation in the form of the expansion in power series in $h$ in a
vicinity of $\mathcal{E}_{0}$. The equation for any state has a pair of such
roots$.$ Notice that the parameter $h\ $is always very small, $%
h=eH/kmc^{2}\ll 1$.

Broadly speaking, parameter $\tilde{p}$ also may depend on $h$. In this case
the eigenvalue equation would bind coefficients of this expansion $\mathcal{E%
}_{k}$ and $\tilde{p}_{k}$ and one from these coefficients, except $\mathcal{%
E}_{0}$ and $\tilde{p}_{0}$, remains indefinite. The possibility to define
both the coefficients is discussed in the last Section. Here we assume that $%
\tilde{p}$ do not depend on $h$.

For the ground state%
\[
\mathcal{E}(\mathcal{E}+\Lambda )-1-\frac{\mathcal{E}h^{2}}{\mathcal{E}-%
\mathcal{E}_{0}}=0,\text{ \ }\Lambda =\frac{2\varepsilon \tilde{p}c-\hbar
\Omega }{mc^{2}}, 
\]

\begin{equation}
\mathcal{E=E}_{0}\pm \frac{\mathcal{E}_{0}h}{\sqrt{\mathcal{E}_{0}^{2}+1}}+%
\frac{\mathcal{E}_{0}h^{2}}{2(\mathcal{E}_{0}^{2}+1)^{2}}+\ldots  \label{Exg}
\end{equation}%
A necessary condition of such an expansion is a certain momentum. This
momentum for the ground state is%
\begin{equation}
\tilde{p}=\frac{\varepsilon mc}{2\mathcal{E}_{0}}-\frac{\varepsilon mc}{2}%
\mathcal{E}_{0}+\varepsilon \frac{\hbar \Omega }{2c}.  \label{pg}
\end{equation}

For the exited state of the first type the expansion coincides with (\ref%
{Exg}), but the term $\varepsilon \hbar \Omega /c$ must be added to momentum.

For the exited state of the second type this expansion is%
\begin{equation}
\mathcal{E}\mathcal{=}\mathcal{E}_{0}\pm \frac{\mathcal{E}_{0}h}{\sqrt{%
\mathcal{E}_{0}^{2}+1+\varsigma }}+\frac{\mathcal{E}_{0}(1+\varsigma )h^{2}}{%
2(\mathcal{E}_{0}^{2}+1+\varsigma )}+\ldots ,\text{ }  \label{eta}
\end{equation}%
where \ $\varsigma =2\hbar \Omega \mathcal{E}_{0}/mc^{2}\ll 1$. The momentum
is the same as for the ground state (\ref{pg}).

\subsection{Average values of operators for single states}

Below we consider average energy, momentum and spin, defined as the integral
of corresponding operators over all the cross-section. This integral has
exact value, however, for simplicity, we use first approximation, neglecting
terms with $h$ and of the order of $\hbar \Omega \ll mc^{2}$. In this
approximation the average value of energy coincides for all considered
states 
\begin{equation}
E_{a}\approx \frac{(\mathcal{E}_{0}^{2}+1)}{\mathcal{E}_{0}}mc^{2}.
\label{Ea}
\end{equation}%
For different states this value differs by terms of the order $\hbar \Omega $%
. In particular, these terms are $+\hbar \Omega /2,$ $-\hbar \Omega
/2,3\hbar \Omega /2$ for the ground, first and second excited states,
respectively.

A minimum of $E_{a}=2mc$ is realized at $\mathcal{E}_{0}=1.$ Noteworthy the
astonishing thing. It may be shown with help of (\ref{Ng}), that this
equality is the classical condition of the magnetic resonance 
\begin{equation}
\mu H_{z}=\frac{1}{2}\hbar \Omega ,  \label{mm}
\end{equation}%
where $\mu =e\hbar /2mc$ is the electron magnetic moment when $g-$factor
equals two. In the general case the inverse of $\mathcal{E}_{0}$ defines the 
$g-$ factor.

The existence of such a minimum allows us to assume that the states have
chances to be stable.

Average components of momenta equal in the first approximation 
\begin{eqnarray}
p_{1a} &=&\mp \frac{1}{2}mc\sqrt{\mathcal{E}_{0}^{2}+1}\cos (\Omega t-kz),
\label{pag} \\
p_{2a} &=&\mp \frac{1}{2}mc\sqrt{\mathcal{E}_{0}^{2}+1}\sin (\Omega t-kz),
\label{pae1} \\
p_{3a} &=&\frac{\varepsilon mc}{\mathcal{E}_{0}}.  \label{pae2}
\end{eqnarray}

Large values of energy and components of average momenta once again can
explain why the term "relativistic" is used for considered solutions.

The average spin is defined by the integral

\begin{equation}
\text{\ }s_{k}=\frac{\hbar }{2}\int \Psi ^{\ast }\sigma _{k}\Psi dxdy,
\label{sk}
\end{equation}%
where $\sigma _{1}=-i\alpha _{2}\alpha _{3,}$ $\sigma _{2}=-i\alpha
_{3}\alpha _{1,}$ $\sigma _{3}=-i\alpha _{1}\alpha $ and the integration is
over all cross-section.

A distinctive feature of singular solutions is the vanishing of $s_{3}$ in
the first approximation. For single states considered here the temporal
behavior of transverse components are 
\begin{eqnarray}
s_{2} &=&\mp \frac{\hbar }{2}\frac{\varepsilon \mathcal{E}_{0}}{\sqrt{%
\mathcal{E}_{0}^{2}+1}}\cos (\Omega t-kz),  \label{sr1} \\
s_{2} &=&\mp \frac{\hbar }{2}\frac{\varepsilon \mathcal{E}_{0}}{\sqrt{%
\mathcal{E}_{0}^{2}+1}}\sin (\Omega t-kz),  \label{sr2}
\end{eqnarray}

\subsection{Mixed states}

Interesting feature of the singular solutions is the existence of different
states with the same momentum $\tilde{p}$, as, for example, the ground and
exited state of the second type. It opens possibility of non-stationary, but
stable mixed states. Such states may be formed only from solutions with the
same sign of the first term in the expression $\mathcal{E}$. For different
signs the integral of the average value of an operator $P$ contains the
complimentary term $\int \Psi _{m}^{\ast }P\Psi _{n}dxdy$, where $m$ and $n$
pertains to different states. This term contains a factor 
\begin{equation}
\exp [-\frac{1}{2d}(d_{2}^{\prime }-d_{2}^{\prime \prime })^{2}]\approx \exp
[-\frac{(\mathcal{E}_{0}^{2}+1)}{2\bar{\lambda}_{e}^{2}d}],  \label{f1}
\end{equation}%
where $\lambda _{e}=3.86\cdot 10^{-11}cm$ is the Compton wavelength of
electron. This factor is very small because $\lambda _{e}^{2}d\ll 1.$ In the
opposite case of the same signs this factor is little differ from 1.

A mixed state consisting of the ground and exited state $\Psi =C_{g}\Psi
_{g}+C_{e2}\Psi _{e2},$ is considered below, as an example. Indexes $g$ and $%
e2$ correspond to the ground and the excite state of the second type
respectively. For simplicity, we assume that $C_{g},C_{e2}$ are real
constants $C_{g}=C_{e2}=1/\sqrt{2}$. For such a mixed state the equality of
momenta is the favorable fact from the viewpoint of stability.

For this state the temporal dependence of spin components in the first
approximation are $s_{3}=0,$%
\begin{eqnarray}
\text{\ }s_{1} &=&\mp \frac{\varepsilon \hbar \mathcal{E}_{0}}{2\sqrt{%
\mathcal{E}_{0}^{2}+1}}[1\pm \cos \omega t]\cos (\Omega t-kz),  \label{sm1}
\\
s_{2} &=&\mp \frac{\varepsilon \hbar \mathcal{E}_{0}}{2\sqrt{\mathcal{E}%
_{0}^{2}+1}}[1\pm \cos \omega t]\sin (\Omega t-kz),  \label{sm2}
\end{eqnarray}%
\begin{equation}
\omega =\frac{\mathit{2}\mathcal{E}_{0}^{2}\omega _{m}}{\sqrt{(\mathcal{E}%
_{0}^{2}+1)^{3}}},\text{ \ }\omega _{m}=\frac{\mu H}{\hbar }.  \label{om}
\end{equation}%
If $C_{g}=-C_{e2}=1/\sqrt{2}$ in (\ref{sm1}), (\ref{sm2}) \ $\cos \omega t$
should be replaced by $\sin \omega t$.\ 

Noteworthy, for comparison, the temporal dependence of the average spin in
the non-relativistic case. It follows from any bounded and square integrable
solutions of the Pauli equation in a rotating magnetic field and arbitrary
static electric field, that this dependence is $s_{1}=(\hbar /2)\sin \omega
_{m}t\sin \Omega t,$ \ $s_{2}=(\hbar /2)\sin \omega _{m}t\cos \Omega t,$ \ $%
s_{3}=(\hbar /2)\cos \omega _{m}t.$ In given case the average spin is
determined by (\ref{sk}), but the integration is over all volume of the
field, $\sigma _{k}$ is the Pauli matrix. .

\section{Non-Galilean transformation for rotating frames of reference}

All the above results are obtained with help of the transition to a rotating
frame. For such a transition the Galilean transformation is used (\ref{xy}),
(\ref{xyt}). It means that time in both frames is assumed to be invariable.
However, from the viewpoint of contemporary physics this assumption seems
unbelievable. This is a strong argument for using a non-Galilean
transformation.

In this Section a general form of normalized non-Galilean transformation 
\begin{eqnarray}
\tilde{r} &=&r,\text{ \ }\tilde{\varphi}=\varphi -\Omega t+kz,\text{ }
\label{Nr} \\
\tilde{t} &=&-\tau \varphi +\gamma z+t,\text{ \ }\tilde{z}=\lambda \varphi
+z+vt,  \label{Nzt}
\end{eqnarray}%
is considered. Here $v$ is the velocity of fermion, $\tau ,$ $\lambda $ and $%
\gamma $ are parameters with the dimension of time, length and inverse
velocity, respectively, these parameters depend of $\Omega $ and $v$.

This linear transformation corresponds to the concept of \ "point rotation
frame" with the axis of rotation at every point \cite{S2}. This concept is
used in optics more 100 years. The characteristic example of the concept is
the optical indicatrix (index ellipsoid).

Normalization of $\tilde{\varphi}$ in (\ref{Nr}) implies that the wave
function in the rotating frame is periodic with respect to $\tilde{\varphi}$
as well as the wave function in the initial frame is periodic with respect
to $\varphi $.

For stationary states in rotating frame the wave function differs from (\ref%
{Psi}) by factor $\exp (-iE\tilde{t}/\hbar +ip\tilde{z}/\hbar -in\tilde{%
\varphi}),$ which is inserted instead of $(-i\tilde{E}\tilde{t}/\hbar +i%
\tilde{p}\tilde{z}/\hbar )$. However, a necessary condition for the
existence of bounded and square integrable solutions of the Dirac equation
must be added 
\begin{equation}
\tau E+\lambda p=\hbar n,  \label{n}
\end{equation}%
where $n$ is a integer. States in the given case are characterized by two
integers similarly to a two-dimensional harmonic oscillator. These integers
are $n$ and the degree of polynomial $\psi $. This is yet another argument
in favor of the non-Galilean transformation.

The condition (\ref{n}) and the eigenvalue equation allows to determine both
parameters $E$ and $p$ in the case dependence them from $h$. This is one
more argument in the favor of this transformation.

Introduce new parameters $\tilde{E}^{\prime }$ and $\tilde{p}^{\prime }$ 
\begin{equation}
\tilde{E}^{\prime }=E-vp-n\hbar \Omega ,\text{ }\ \tilde{p}^{\prime }=-\frac{%
1}{v_{z}}E+p-\frac{\varepsilon }{c}n\hbar \Omega .  \label{Eprime}
\end{equation}%
$\tilde{E}^{\prime }$ and $\tilde{p}^{\prime }$ coincide with $\tilde{E}$
and $\tilde{p},$ only in a first approximation. With these parameters and
due to condition (\ref{n}), the form of the equation for stationary states
in both cases of the Galilean and non-Galilean transformation fully
coincides. Using (\ref{Eprime}) $E$ and $p$ may be expressed in terms of $%
\tilde{E}^{\prime }$ and $\tilde{p}^{\prime }$ and inserted in (\ref{n})$.$

It may be straightforwardly shown with help of the condition (\ref{n}) and
relations (\ref{Nr}), (\ref{Nzt}) that in the initial frame the equality 
\begin{equation}
(-i\frac{E\tilde{t}}{\hbar }+i\frac{p\tilde{z}}{\hbar }-in\tilde{\varphi}%
)=(-i\frac{\tilde{E}t}{\hbar }+i\frac{\tilde{p}z}{\hbar })  \label{Ep}
\end{equation}%
is fulfilled. In this frame the shape of wave functions for both cases
coincide since for the Galilean transformation $\tilde{t}=t,\tilde{z}=z$.

In the general case, for arbitrary $n$, all expressions as well as the
condition (\ref{n}) are too cumbrous therefore we present here this
condition for $n=0,$ as an illustration%
\begin{equation}
(\tau +\frac{\lambda }{v_{z}})(\frac{1}{\mathcal{E}_{0}}+\mathcal{E}%
_{0}+\eta )+(\tau v+\lambda )\frac{\varepsilon }{c}(\frac{1}{\mathcal{E}_{0}}%
-\mathcal{E}_{0}+\eta )=0.  \label{con0}
\end{equation}%
Another parameter, which may be measured, is the frequency $\omega $. This
frequency differs from (\ref{om}), and at $n=0$ is as follows 
\begin{equation}
\omega =\frac{(1+\mathcal{E}_{0}^{2})}{2\sqrt{2}}\omega _{m}.  \label{om1}
\end{equation}%
Once again the surprising fact, both frequencies coincide provided $\mathcal{%
E}_{0}^{2}=1.$

\section{Conclusion}

Fermions in the field of a traveling, circularly polarized electromagnetic
wave and a constant magnetic field are localized in the small cross-section
with the size of the order of $l_{c}$. Singular solutions, i.e., solutions
with energy in vicinity of the singular point are of considerable interest,
from viewpoint of experiment. Such solutions arise only at certain values of
the longitudinal momentum. The average energy for singular states may exceed
more than twice the energy of rest mass. In contrast to the non-relativistic
case, the longitudinal component of spin for all states equals zero in the
first approximation.

For single states the transverse momentum and spin are oscillated similarly
to the amplitude magnetic or electric field of the traveling circularly
polarized wave. For mixed states temporal behavior of transverse components
of spin differs from the non-relativistic case by the form of the temporal
change. Average energy of such states has a minimum of the order of $2mc^{2}$%
. The condition of this minimum exactly coincides with the classical
condition of magnetic resonance at $g$-factor 2. Possible, mixed states
consisting of states with the same momentum, at the minimum energy, have
chances to be stable.

Physical arguments in favor of the statement that the transformation, for
the transition to the rotating frame, must be non-Galilean are presented in
the paper.

From the above studies the conclusion follows that some parameters of this
transformation can be measured not only in optics, but also in high-energy
physics. Such measurements are very important since the parameters determine
limits of using physical laws on very small intervals of time $<10^{-23}\sec 
$ and length $<10^{-13}cm$.

\end{document}